\documentclass[12pt]{article}

\begin{document}

\begin{titlepage}

\begin{flushright}
IUHET-485\\
hep-th/0507258
\end{flushright}
\vskip 2.5cm

\begin{center}
{\Large \bf Lorentz Violation and Synchrotron Radiation}
\end{center}

\vspace{1ex}

\begin{center}
{\large B. Altschul\footnote{{\tt baltschu@indiana.edu}}}

\vspace{5mm}
{\sl Department of Physics} \\
{\sl Indiana University} \\
{\sl Bloomington, IN 47405 USA} \\

\end{center}

\vspace{2.5ex}

\medskip

\centerline {\bf Abstract}

\bigskip

We consider the radiation emitted by an ultrarelativistic charged
particle moving in a
magnetic field, in the presence of an additional
Lorentz-violating interaction. In contrast with prior work, we treat a
form of Lorentz violation that is represented by a renormalizable operator.
Neglecting the radiative reaction force, the particle's
trajectory can be determined exactly.
The resulting orbit is generally noncircular and does not lie in the plane
perpendicular to the magnetic field. We
do not consider any Lorentz violation in the electromagnetic sector, so the
radiation from the accelerated charge can be determined by standard means, and
the radiation spectrum will exhibit a
Lorentz-violating directional dependence. Using data on
emission from the Crab nebula, we can set a bound on a particular combination
of Lorentz-violating coefficients at the $6\times10^{-20}$ level.

\bigskip

\end{titlepage}

\newpage

\section{Introduction}

There is presently a great deal of interest in the possibility that
Lorentz and CPT invariance may be violated in nature. If the fundamental laws
of physics do not obey these symmetries, then we would expect to see evidence of
this
violation even in the low-energy effective theory. Therefore, if small Lorentz or
CPT violations were discovered, they would represent crucial clues about
the structure of the most basic theory of nature. A general standard model
extension (SME), containing possible Lorentz- and CPT-violating corrections to
quantum field theory~\cite{ref-kost1,ref-kost2} and general
relativity~\cite{ref-kost12} has been developed. The SME offers a parameterization
of Lorentz violations in low-energy effective field theory, and both its
renormalizability~\cite{ref-kost4} and stability~\cite{ref-kost3} have been
carefully examined.

The SME provides a theoretical framework for analyzing experimental results.
Sensitive tests of Lorentz symmetry have included studies of matter-antimatter
asymmetries for
trapped charged particles~\cite{ref-bluhm1,ref-bluhm2,ref-gabirelse,
ref-dehmelt1} and bound state systems~\cite{ref-bluhm3,ref-phillips},
determinations of muon properties~\cite{ref-kost8,ref-hughes}, analyses of
the behavior of spin-polarized matter~\cite{ref-kost9,ref-heckel},
frequency standard comparisons~\cite{ref-berglund,ref-kost6,ref-bear},
measurements of neutral meson
oscillations~\cite{ref-kost10,ref-kost7,ref-hsiung,ref-abe},
polarization measurements on the light from distant galaxies~\cite{ref-carroll1,
ref-carroll2,ref-kost11}, and others. The results of these experiments can be
used to set bounds on various SME coefficients. Many coefficients are very
strongly constrained, but many others are not.

There are many systems and reaction processes that could potentially be used to
set further bounds of the SME coefficients for Lorentz violation.
We shall consider a particular process---synchrotron motion and radiation---and
examine how it would be impacted by Lorentz violation. Although there have
been many
analyses of this process in the presence of Lorentz-violating dispersion
relations, there is as yet no analysis in terms of the renormalizable operators
of the SME.

Analyses of possible Lorentz violation in synchrotron emission have often
focused only on changes to particle dispersion relations. One popular approach is
that of Myers and Pospelov~\cite{ref-myers}. Taking a preferred direction
$v^{\mu}$ in spacetime, one may add an operator proportional to $i\phi^{*}\left(
v^{\mu}\partial_{\mu}\right)^{3}\phi$ to the Lagrange density for a scalar
particle. If $v^{\mu}$ has a time
component only, this will add a term proportional to $E^{3}$
to the usual relativistic energy-momentum relation $E^{2}=\vec{p}\,^{2}+m^{2}$.
Of course, the statement that $v^{\mu}$ is purely timelike is not Lorentz
invariant, so that assumption must be taken to hold is some particular preferred
frame, which is typically the rest frame of the cosmic microwave background. The
electromagnetic field is incorporated through the usual minimal coupling
procedure. In the presence of this kind of Lorentz violation, the motion of a
charged particle in a constant magnetic field is modified, but the projection
of the trajectory onto
the plane perpendicular to $\vec{B}$ remains circular, and the particle's speed
remains constant. The radiation in the far field can be determined, including
information about polarization, and circumstances that could enhance observable
effects have been identified~\cite{ref-motemayor1,ref-motemayor2}.

Stringent bounds on Lorentz violations with modified dispersion
relations have been obtained from data from the Crab nebula~\cite{ref-jacobson1,
ref-jacobson2,ref-ellis}. These modifications
can lead to maximum particle velocities that are less than the speed of light,
but the Crab nebula shows evidence of synchrotron emission from electrons with
Lorentz factors of $\gamma=\left(1-\vec{v}\,^{2}\right)^{-1/2}\sim3\times10^{9}$,
or energies of 1500 TeV. So the
existence of electrons with velocities this large can be used to constrain
models with deformed dispersion relations. For a Lorentz-violating coefficient
with a particular sign, the data show that the coefficient must be at least
seven orders of magnitude smaller than ${\cal O}(E/M_{P})$ Planck-level
suppression.

Lorentz violation can also be incorporated into particle physics through the
introduction of noncommutative field theory~\cite{ref-carroll3}. Synchrotron
radiation has also been analyzed within this framework. The minimal coupling
between charged matter and the electromagnetic field is modified by the
noncommutativity, as is the structure of the free radiation field itself.
A discussion
in~\cite{ref-castorina} focuses on the particular case in which the magnetic
field and the Lorentz-violating noncommutativity parameter are aligned, so that
the orbits of charged particles in the plane perpendicular to $\vec{B}$ are
again given by circles. It is possible to work out the far fields within this
model, at leading order in the noncommutativity, but there are a number a
difficulties, including acausality and potential problems with quantization.

However, these analyses ignore some of the most natural
Lorentz-violating operators.
There is a unique spin-independent, superficially renormalizable
SME coupling that is consistent with the
gauge invariance of the standard model and which grows in relative
importance at high
energies. This is a CPT-even two-index tensor $c^{\nu\mu}$. We shall look at
how the presence of such a constant background tensor (which could arise, for
example, as the vacuum expectation value of a dynamical tensor field) will
modify synchrotron emission. Using existing data about the nonthermal spectrum of
the Crab nebula, we may place a bound of $6\times10^{-20}$ on a particular
linear combination of the $c^{\nu\mu}$ coefficients. The method by which we find
this
bound is very similar to that used to bound other types of Lorentz violation;
however, the $c^{\nu\mu}$ interaction is actually much more natural to consider
than these,
because it is superficially renormalizable.
All the analyses so far have been essentially classical in nature, and
we shall continue working within the classical framework, although we shall
look at when quantum corrections would become important.

\section{Synchrotron Motion with Lorentz Violation}

To study synchrotron motion,
we shall consider a theory of fermions interacting with the electromagnetic
field. The Lagrange density for this theory is
\begin{eqnarray}
{\cal L} & = & -\frac{1}{4}F^{\mu\nu}F_{\mu\nu}+
\bar{\psi}[\Gamma^{\mu}(i\partial_{\mu}-eA_{\mu})-m]\psi \nonumber\\
\label{eq-L}
& = & -\frac{1}{4}F^{\mu\nu}F_{\mu\nu}+
\bar{\psi}[(\gamma^{\mu}+c^{\nu\mu}\gamma_{\nu})
(i\partial_{\mu}-eA_{\mu})-m]\psi.
\end{eqnarray}
The $c^{\nu\mu}$ interaction is the source of the Lorentz violation. There are
other superficially renormalizable couplings contained in the Standard Model
Extension, but the $c$ couplings are most natural in this context. When
considering synchrotron radiation, one is primarily interested in particles
with very high energies. Lorentz-violating coefficients that modify the kinetic
part of the Lagrangian will grow in relative
importance at high energies, so it is natural
to consider only these kinetic modifications. There are only two such sets of
Lorentz-violating terms that are consistent with the more general standard model
gauge couplings---the $c$ terms and also
a set of $d^{\nu\mu}$ terms, which have the same form as the $c$ interactions,
except for the addition of a $\gamma_{5}$.

However, we shall not consider the $d$ interactions
here. They are spin-dependent, while the $c$ term exists for bosonic
(Klein-Gordon) particles as well as fermions. So all our results will apply
equally to the motion of spin-zero charged particles. Moreover,
spin precession effects will naturally decrease the importance of any $d$ terms.
For an electron undergoing circular cyclotron motion, with the spin oriented in
the plane of the orbit, the spin rotates by
$2\pi\gamma\frac{g-2}{2}$ radians with each orbital revolution. For
$\gamma\gg\alpha^{-1}$, the spin will rotate many times during one orbital
period, and any effects proportional to the helicity will be diminished by the
resultant averaging.

Modifications of the kinetic Lagrangian that are not invariant under the standard
model's $SU(2)_{L}$ gauge symmetry can also exist; however, they can only appear
as part of electroweak symmetry breaking, as vacuum expectation values of
nonrenormalizable operators. These operators should therefore be further
suppressed, and we shall neglect them.

We shall also neglect any Lorentz
violation in the photon sector. Modifications of the free electromagnetic
Lagrangian will generally change the speed of photon propagation. This leads to
the possibility of vacuum Cerenkov
radiation~\cite{ref-lehnert1,ref-lehnert2}, which is not yet fully understood,
although threshold analyses can be used to set further limits on
Lorentz-violating parameters. Most possible Lorentz-violating terms in the free
electromagnetic sector also give rise to photon birefringence, which has been
searched for and not seen. The limits on the
relevant forms of Lorentz violation are very
strong, and we may safely neglect them. The purely electromagnetic
terms that do not cause birefringence can be accounted for by adding 
\begin{equation}
{\cal L}_{F}=-\frac{1}{4}\left(k_{F}\right)^{\alpha}\, _{\mu\alpha\nu}
\left(F^{\rho\mu}F_{\rho}\,^{\nu}+F^{\mu\rho}F^{\nu}\, _{\rho}\right).
\end{equation}
to ${\cal L}$. However, a coordinate transformation $x^{\mu}\rightarrow x^{\mu}
-\frac{1}{2}\left(k_{F}\right)^{\alpha\mu}\,_{\alpha\nu}x^{\nu}$ will eliminate
all the Lorentz violation from the photon sector at leading
order~\cite{ref-kost16,ref-kost17}. This transformation shifts
the Lorentz-violating physics into the charged matter sector, where it
manifests itself exactly as a $c^{\nu\mu}$ term. We see that consideration of $c$
therefore captures all the possible sources of Lorentz violation in a
synchrotron process that are not significantly further suppressed.
However, the transformation that eliminates $k_{F}$ is
frame-dependent, and the new coordinates need not even be rectangular relative to
the original ones; so by choosing to consider only this form of Lorentz violation,
we are restricting ourselves to working in a very particular and special
coordinate system.

We know that the Lorentz-violating coefficients for any physical charged
particles are small. Physically, we might expect that the characteristic size
for $c^{\nu\mu}$ is ${\cal O}(m/M_{P})$. However, we shall not make any
special use of the fact that $M_{P}$ is the Planck scale. Rather, we may take
this size estimate as effectively defining $M_{P}$; $M_{P}$ is whatever large
energy scale is needed in order to give the $c$ terms the correct magnitude.

The canonical quantization of the fermion field requires some care when the
$c$ coefficients are nonvanishing. If $c^{\nu0}$ is nonzero, then
${\cal L}$ will contain nonstandard time derivative terms.
In this case, a matrix transformation $\psi\rightarrow R\psi$ will be required,
to ensure that $\Gamma^{0}=\gamma^{0}$. An explicit power-series expression for
the required $R$ is given in~\cite{ref-lehnert3}.
For simplicity, we shall assume that any such necessary
transformation has already been performed and $c^{\nu0}=0$.
However, this will require us to consider the canonical quantization in a single
frame only. We may not boost the theory into another frame, because doing so
would reintroduce the problematic time derivatives.

In fact, in much of what follows, we shall neglect the $c^{0\mu}$ terms as well.
While there is no special reason to believe this, we shall assume that the
Lorentz violation is purely spacelike in a frame in which $F^{\mu\nu}$ contains
only a magnetic component. We do this because the problem can then be solved
exactly, to all orders in the remaining Lorentz-violating coefficients.
However, when we revert to the linearized approximation and derive a limit on
the $c$ coefficients from the observed properties of the synchrotron spectrum,
we shall include the $c^{0\mu}$ parts in the calculation.

We shall consider the interaction with the electromagnetic field in two stages.
This is standard practice in consideration of cyclotron motion. First, we
determine the path traced out by a nonradiating charged particle moving in a
spatially homogeneous
background magnetic field. Then we evaluate the radiation induced by this
periodic motion.

Because we are interested in the synchrotron emission from a single particle,
we shall make the natural approximation of treating the Dirac equation as a
single-particle wave equation. Standard techniques of relativistic quantum
mechanics then apply; however, Lorentz violation will introduce new
complexities.
The momentum and velocity are not simply related by $\vec{v}=\vec{\pi}/E=
\vec{\pi}/\gamma m$. The single-particle fermion Hamiltonian derived from
${\cal L}$ is
\begin{equation}
H=\alpha_{j}\pi_{j}-c_{lj}\alpha_{l}\pi_{j}-c_{0j}\pi_{j}+\beta m,
\end{equation}
where $\alpha_{j}=\gamma^{0}\gamma^{j}$ and $\beta=\gamma^{0}$ are the usual
Dirac matrices, and $\vec{\pi}$ is the mechanical (rather than canonical)
three-momentum $\vec{\pi}=\vec{p}-e\vec{A}$. Time derivatives of
operators relating to fermion properties
may be found by taking commutators with this Hamiltonian. In particular, the
derivative of the particle's position is
\begin{equation}
\label{eq-xdot}
\dot{x}_{k}=\alpha_{k}-c_{lk}\alpha_{l}-c_{0k}.
\end{equation}
Similarly, the equation of motion for $\alpha_{k}$ may be written
\begin{equation}
\dot{\alpha}_{k}=i\left[-2\alpha_{k}(H+c_{0j}\pi_{j})+2\pi_{k}-2c_{kj}\pi_{j}
\right],
\end{equation}
which has the exact solution
\begin{equation}
\label{eq-alphasol}
\alpha_{k}(t)=\left(\pi_{k}-c_{kj}\pi_{j}\right)(H+c_{0j}\pi_{j})^{-1}+\left[
\alpha_{k}(0)-\left(\pi_{k}-c_{kj}\pi_{j}\right)(H+c_{0j}\pi_{j})^{-1}\right]
e^{-2i(H+c_{0j}\pi_{j})t}.
\end{equation}

The second term on the right-hand-side of (\ref{eq-alphasol}) is matrix-valued
and oscillatory. This term describes the particle's {\em Zitterbewegung}, which,
for a well-localized wave packet, is generated by interference between positive
and negative frequency (i.e. particle and antiparticle) modes. The first term,
when combined with (\ref{eq-xdot}) gives the bulk velocity~\cite{ref-altschul4}
\begin{equation}
\label{eq-vfull}
v_{k}=\frac{1}{E+c_{0j}\pi_{j}}\left(\pi_{k}-c_{kj}\pi_{j}-c_{jk}\pi_{j}+c_{jk}
c_{jl}\pi_{l}\right)-c_{0k}.
\end{equation}
This same expression can also be found by calculating the group velocity
$\vec{v}_{g}=\vec{\nabla}_{\vec{\pi}}E$. However, as previously stated, we will
drop the $c_{0j}$ contributions in much of the following and use
\begin{equation}
\label{eq-v}
v_{k}=\frac{1}{E}\left(\pi_{k}-c_{kj}\pi_{j}-c_{jk}\pi_{j}+c_{jk}c_{jl}\pi_{l}
\right).
\end{equation}

If rotation invariance is unbroken in the inertial frame we are considering, so
that $c_{jk}\propto\delta_{jk}$, there is merely a rescaling of the velocity.
This will lead to fewer interesting effects. Fortunately, even if there is a
privileged frame in which $c_{jk}\propto\delta_{jk}$, a body emitting
synchrotron radiation will not generally be at rest in this frame. We shall
therefore assume that there is some breaking of rotation invariance in the rest
frame of the source. In general, we shall assume that there is no suppression of
rotation invariance violation relative to boost invariance violation only.

The equation of motion for the particle is the unmodified Lorentz force law
$\dot{\vec{\pi}}=e\dot{\vec{x}}\times\vec{B}$.
We shall neglect the {\em Zitterbewegung} in $\dot{\vec{x}}$ and
consider only
\begin{equation}
\label{eq-pidot}
\dot{\vec{\pi}}=e\vec{v}\times\vec{B}.
\end{equation}
Then, since according to (\ref{eq-v}) $\vec{v}$ remains
a linear function of the momentum, we may solve for the particle's motion
exactly. Again, we emphasize that all these same results for the bulk velocity and
equation of motion also apply to
Klein-Gordon particles, although the Klein-Gordon equation is even less
satisfactory as a single-particle wave equation than is the Dirac equation.

It is advantageous to solve for the time development of the velocity, rather than
the momentum (canonical or mechanical). While the velocity is a less fundamental
object, the greatest formal problem with it---the {\em Zitterbewegung}---has
already been neglected.
The momentum could be determined with equal ease, but it
possesses the unattractive property that a particle could possess zero momentum,
yet not be stationary, if a $c_{0j}$ term were present. Moreover,
even with the Lorentz-violation, the electromagnetic field is coupled directly to
the velocity. Because $c^{\nu0}=0$, the electrostatic potential $\Phi=A^{0}$ is
coupled, as usual, to the charge density $e\psi^{\dag}\psi$. Similarly, the
vector potential $\vec{A}$ couples to $e\psi^{\dag}\dot{\vec{x}}\psi$,
where $\dot{\vec{x}}$ is given by (\ref{eq-xdot}). Neglecting the
{\em Zitterbewegung}, the coupling is simply to the bulk velocity $\vec{v}$.
The fact that that the electromagnetic coupling is standard in this way was
already evident in the Lorentz force law (\ref{eq-pidot}), and it holds equally
in the equations of motion for $A^{\mu}$.

To determine the particle's motion, we must solve a set of two coupled
differential equations. These two equations describe the time evolution of the
two components of the velocity in the plane perpendicular to $\vec{B}$; the
component of $\vec{v}$ parallel to $\vec{B}$ does not contribute to the $\vec{v}
\times\vec{B}$
force. Let us take $\vec{B}$ to point along the $z$-direction, $\vec{B}
=B\hat{z}$. Then the equations of motion for $\vec{\pi}$ are
\begin{eqnarray}
\label{eq-pi1dot}
\dot{\pi}_{1} & = & eBv_{2} \\
\label{eq-pi2dot}
\dot{\pi}_{2} & = & -eBv_{1} \\
\dot{\pi}_{3} & = & 0.
\end{eqnarray}
So $\pi_{3}$ is a constant of the motion, as is $E=\sqrt{m^{2}+\left(\pi_{k}
-c_{kj}\pi_{j}\right)\left(\pi_{k}-c_{kl}\pi_{l}\right)}$.
Differentiating (\ref{eq-v}) then gives the following equations of motion for
$v_{1}$ and $v_{2}$
\begin{eqnarray}
\label{eq-v1dot}
\dot{v}_{1}  & = & \frac{1}{E}\left[\left(1-2c_{11}+c_{j1}c_{j1}\right)
\dot{\pi}_{1}+\left(-c_{12}-c_{21}+c_{j1}c_{j2}\right)\dot{\pi}_{2}\right] \\
\label{eq-v2dot}
\dot{v}_{2}  & = & \frac{1}{E}\left[\left(-c_{12}-c_{21}+c_{j1}c_{j2}\right)
\dot{\pi}_{1}+\left(1-2c_{22}+c_{j2}c_{j2}\right)\dot{\pi}_{2}\right].
\end{eqnarray}
Combining equations (\ref{eq-pi1dot}), (\ref{eq-pi2dot}), (\ref{eq-v1dot}),
and (\ref{eq-v2dot}) in matrix form gives
\begin{equation}
\label{eq-vdot}
\left[
\begin{array}{c}
\dot{v}_{1} \\
\dot{v}_{2}
\end{array}
\right]=\frac{eB}{E}\left[
\begin{array}{cc}
-\beta & \alpha \\
-\gamma & \beta
\end{array}
\right]\left[
\begin{array}{c}
v_{1} \\
v_{2}
\end{array}
\right]=\omega_{0}M\left[
\begin{array}{c}
v_{1} \\
v_{2}
\end{array}
\right].
\end{equation}
The elements of the matrix $M$ are $\alpha=\left(1-2c_{11}+
c_{j1}c_{j1}\right)$, $\beta=\left(-c_{12}-c_{21}+c_{j1}c_{j2}
\right)$, and $\gamma=\left(1-2c_{22}+c_{j2}c_{j2}\right)$, and
$\omega_{0}=\frac{eB}{E}$.

The equation (\ref{eq-vdot}) is easily solved. Since $M^{2}=-(\alpha\gamma-
\beta^{2})I$ (where $I$ is the identity matrix), $e^{\omega_{0}Mt}=I\cos\omega t
+\frac{\omega_{0}}{\omega}M\sin\omega t$, where $\omega=\omega_{0}
\sqrt{\alpha\gamma-\beta^{2}}$.
For vanishing $c^{\nu\mu}$, $\omega=\omega_{0}$ is the usual
synchrotron
frequency. If we choose coordinates so that the initial conditions are
$v_{1}(t=0)=v_{10}$ and $v_{2}(0)=0$, then
\begin{equation}
\label{eq-v12sol}
\left[
\begin{array}{c}
v_{1}(t) \\
v_{2}(t)
\end{array}
\right]=e^{Mt}\left[
\begin{array}{c}
v_{10} \\
0
\end{array}
\right]=v_{10}\left[
\begin{array}{c}
\cos\omega t+\frac{\omega_{0}}{\omega}\left(c_{12}+c_{21}-c_{j1}c_{j2}\right)
\sin\omega t \\
-\frac{\omega_{0}}{\omega}\left(1-2c_{22}+c_{j2}c_{j2}\right)\sin\omega t
\end{array}
\right].
\end{equation}

The velocity in the $z$-direction can be found by direct integration of its
derivative,
\begin{equation}
\dot{v}_{3}=\frac{eB}{E}\left[\left(-c_{13}-c_{31}+c_{j1}c_{j3}\right)v_{2}
+\left(c_{23}+c_{32}-c_{j2}c_{j3}\right)v_{1}\right].
\end{equation}
So, if $v_{3}(0)=v_{30}$,
\begin{eqnarray}
v_{3}(t) & = & v_{10}\frac{\omega_{0}}{\omega}\left\{(c_{23}+c_{32}-c_{j2}c_{j3})
\left[\sin\omega t-\left(\frac{\omega_{0}}{\omega}\right)(c_{12}+c_{21}-c_{j1}
c_{j2})(\cos\omega t -1)\right]\right. \nonumber\\
\label{eq-v3sol}
& & -\left.\left(\frac{\omega_{0}}{\omega}\right)
(c_{13}+c_{31}-c_{j1}c_{j3})(1-2c_{22}+c_{j2}c_{j2})
(\cos\omega t-1)\right\}+v_{30}.
\end{eqnarray}
The particle moves in an elliptical helix; there is a constant drift
parallel to $\vec{B}$, superimposed upon an additional periodic motion.
If the drift vanishes, then the orbit lies close to, but is not generally
in, the plane normal to the magnetic field, because $v_{3}$ does not generally
vanish, even if its mean value does.

\section{Radiation Emission}

We shall now move on to the second stage our calculation.
We have the particle's motion prescribed, so we may study the radiation
emitted during this motion. For
simplicity, we shall consider only the case in which the drift velocity is zero.
[This does not correspond to $v_{30}=0$, because there are additional
time-independent terms in (\ref{eq-v3sol}). Instead, the sum of these
constant terms must vanish.]
However, since we have now formulated the problem in terms of a particle of
prescribed
velocity conventionally coupled to the radiation field, normal boosting
techniques can be used to generalize these results to a situation in which
the time-averaged velocity in the $z$-direction is nonvanishing.
A crucial quantity to calculate is the speed of the
particle, $|\vec{v}\,|$, which is given by
\begin{eqnarray}
\vec{v}\,^{2} & = & \frac{v_{10}^{2}}{2}\left[\left(\eta+\xi\right)+\left(\eta-
\xi\right)\cos2\omega t+\zeta\sin 2\omega t\right] \\
\label{eq-v2}
& = & \frac{v_{10}^{2}}{2}\left[\left(\eta+\xi\right)+\sqrt{(\eta-\xi)^{2}+
\zeta^{2}}\cos\left(2\omega t-2\phi\right)\right],
\end{eqnarray}
where $\tan 2\phi=\zeta/(\xi-\eta)$ and the constants $\eta$, $\xi$, and $\zeta$
are
\begin{eqnarray}
\eta & = & 1 +\left(\frac{\omega_{0}}{\omega}\right)^{4}
[(c_{23}+c_{32}-c_{j2}c_{j3})(c_{12}+c_{21}-c_{j1}c_{j2}) \nonumber\\
& & +(c_{13}+c_{31}-c_{j1}c_{j3})(1-2c_{22}+c_{j2}c_{j2})]^{2} \\
\xi & = & \left(\frac{\omega_{0}}{\omega}\right)^{2}
\left[(c_{12}+c_{21}-c_{j1}c_{j2})^{2}+(1-2c_{22}+c_{j2}c_{j2})^{2}+
(c_{23}+c_{32}-c_{j2}c_{j3})^{2}\right] \\
\zeta & = & 2\frac{\omega_{0}}{\omega}(c_{12}+c_{21}-c_{j1}c_{j2})-2\left(
\frac{\omega_{0}}{\omega}\right)^{3}(c_{23}+c_{32}-c_{j2}c_{j3}) \\
& & \times[(c_{23}+c_{32}-c_{j2}c_{j3})(c_{12}+c_{21}-c_{j1}c_{j2})+
(c_{13}+c_{31}-c_{j1}c_{j3})(1-2c_{22}+c_{j2}c_{j2})]. \nonumber
\end{eqnarray}

Thus far, our results have been exact, except that we have neglected the radiative
reaction force. Henceforth, we shall be making use of the standard,
Lorentz-invariant results on
the power radiated by a particle undergoing synchrotron
motion~\cite{ref-schwinger1}.
However, the standard methodology for evaluating synchrotron
emission involves a number of approximations. One often neglects any effect
suppressed by a positive power of the Lorentz factor $\gamma$, and we shall follow
this prescription. Among the things we may therefore neglect is the
radiation due to the component of the acceleration parallel to the velocity;
this contribution to the emission is small in comparison with that arising from
the perpendicular component of the acceleration.
We may also ignore the angular
width of the radiation beam. All the emitted energy is beamed into a narrow pencil
of angles
centered around the instantaneous direction of the velocity. The range of angles
covered is ${\cal O}(\gamma^{-1})$, but we shall neglect this spread,
instead assuming that all radiation is emitted along a ray tangent to the
particle's path.

We shall also neglect the Lorentz violation as a source of angular deviation.
Although the exact orbit is neither circular nor in the plane normal to $\vec{B}$,
the deviations from the conventional trajectory are small, of
${\cal O}(c)$. It would not be
feasible to measure changes in the angular distribution of the emitted radiation
induced by the presence of
the Lorentz violation. We shall therefore neglect the changes in the
orbital shape. All effects we shall consider will therefore be related to the
modification of $|\vec{v}\,|$ (\ref{eq-v2}). (This is similar to the approach
adopted in~\cite{ref-jacobson1}, where the magnitude of the velocity was also
taken as the central quantity.)
As the velocity changes around the particle's nearly circular path, the rate at
which radiation is emitted will vary. The most sensitive tests of $c$-type Lorentz
violation in synchrotron radiation could come from comparing the power output
in different directions. (Unfortunately, such measurements are obviously not
possible for single astrophysical sources.)

The phase $\phi$ represents the angular position of the particle in its
orbit at the time when the velocity is a maximum. At the antipodal point of the
orbit, the velocity is also maximal. The greatest amount of radiation is then
emitted along the tangent rays at these two points and propagates in the
directions given by the azimuthal angles $\phi\pm\frac{\pi}{2}$. Similarly, the
smallest radiated power is in the directions $\phi$ and $\phi+\pi$. The presence
of this
effect is of course dependent on the existence of rotation invariance violation.

Neglecting radiation due to the component of the acceleration parallel to the
velocity [which is smaller by a factor of ${\cal O}(\gamma^{-2})$],
the intensity spectrum per unit spectral frequency $\omega_{s}$ is
\begin{equation}
\frac{dI}{d\omega_{s}}=\sqrt{3}e^{2}\gamma\frac{\omega_{s}}{\omega_{c}}
\int_{\omega_{s}/\omega_{c}}^{\infty}dx\, K_{5/3}(x).
\end{equation}
The critical frequency is $\omega_{c}=\frac{3}{2}\gamma^{3}\rho^{-1}$, and
$\rho$ is the instantaneous radius of curvature of the orbit $\rho=\vec{v}\,^{2}
/|\vec{a}^{\perp}|$, where $\vec{a}^{\perp}$ is the component of the acceleration
perpendicular to $\vec{v}$, $\vec{a}^{\perp}=\dot{\vec{v}}-\frac{\dot{\vec{v}}
\cdot\vec{v}}{\vec{v}\,^{2}}\vec{v}$. Neglecting the Lorentz-violating
corrections, $\rho$ is approximately $E/|e|B$.
$K_{5/3}(x)$ is a modified Bessel function of the second kind.
One could go further and calculate the radiation fields in the far field
explicitly. However, we shall not do this, because the polarization structure
of the emitted radiation is not substantially effected by the Lorentz violation.

For ultrarelativistic particles, for which $1-|\vec{v}\,|\ll 1$, the 
Lorentz factor is roughly $\gamma\approx1/\sqrt{2(1-|\vec{v}\,|)}$, and this is
a rapidly increasing function of the speed---$d\gamma/d|\vec{v}\,|=|\vec{v}\,|
\gamma^{3}\approx\gamma^{3}$. The description of the Lorentz violation through
an effective field theory containing only $c^{\nu\mu}$ terms will break down if
the modifications of the velocity due to the presence of $c$ can render the
speed superluminal. According to (\ref{eq-v}), this can occur when $|\vec{\pi}|/E
\approx 1-|c|$, where $|c|$ is a characteristic size for the Lorentz-violating
coefficients. This gives us an estimate of the maximum value of $\gamma$ that
can be achieved before new physics must come into play if some form of causality
is to be preserved: $\gamma_{\max}\sim1/\sqrt{|c|}$. This corresponds to an
energy scale $E_{\max}\sim\sqrt{mM_{P}}$.

\section{Prospects for Observability}

It is still unclear whether the changes we have described in the emission
will be observable, and we shall now turn our attention to this issue.
The total radiated synchrotron power in the ultrarelativistic regime is
proportional to
$\gamma^{4}$. Therefore, the change in the radiated power as the velocity
varies around the orbit is given by
\begin{equation}
\label{eq-DeltaP}
\frac{\Delta P}{P}\approx 2\frac{\left[d\left(\gamma^{4}\right)/d|\vec{v}|\right]
|c|}{\gamma^{4}}\approx 8\gamma^{2}|c|.
\end{equation}
The factor of 2 comes from the fact that the deviations in $|\vec{v}\,|$ range
over both positive and negative values.
The characteristic size $|c|$ used in this calculation should
be essentially the same as that used in the determination of $\gamma_{\max}$,
because in both cases, $|c|$ measures the
magnitude of the contribution that $c^{\nu\mu}$ can make to the velocity.
Inserting $\gamma=\gamma_{\max}$ into (\ref{eq-DeltaP}) gives a result that is
greater than unity. This means that the fractional change in the emitted power
can be of order one in the regime in which the theory is valid; we do not have
to go to an energy scale so high that new physics must emerge in order to see
changes in the emission. However, we do need to get comparatively close to
the scale at which the theory breaks down in order to observe
deviations in the spectrum.

In fact, for $|c|\approx10^{-19}$, we will find a $\frac{\Delta P}{P}$ of one
percent at $\gamma\approx10^{8}$. For the lightest charged particle, the electron,
this corresponds to an energy of roughly 50 TeV, orders of magnitude beyond
anything one could create in the laboratory. We conclude that these effects are
unobservable for Earth-based sources.

The only sources of synchrotron radiation that are high enough in energy to
give observable results of the type we are considering are astrophysical. However,
as each astrophysical source can only be observed from a single direction,
more than one source would be required in order to make the kind of directional
observations that could constrain $c$ most strongly. Ideally, we would want to
have two or more
very clean sources of synchrotron radiation, for which the spectra due to the
motion of multiple species of particles (e.g., both electrons and protons) could
be resolved. Then we could look for systematic differences between the emission
profiles for the species. This would mean effectively using the proton
spectra, for example, as local magnetometers and looking to see whether the
electron spectra are consistent with the measured fields. One could then set
bounds on a combination of the $c$ coefficients for the electron and the proton.

Although observations of distant synchrotron sources (such as far-off radio
galaxies) are ideal for constraining Lorentz violation in the photon sector, they
are not so helpful here. A long line of sight will magnify small effects that
modify the propagation structure of radiation. However, large distances do nothing
to assist measurements of Lorentz violation in the charged emitters themselves. A
nearby, accurately understood source is better than a distant one.

The best-understood synchrotron source is the Crab nebula, but its spectrum
still appears too complicated for the kind of procedure
we have suggested to be at all feasible.
(For a good review of the Crab nebula's nonthermal emission spectrum,
see~\cite{ref-aharonian}.)
For example, the spectrum contains two different
electron synchrotron components, with significantly different characteristics.
Any observed proton synchrotron radiation would probably fail as a sensitive
magnetometer, because it would be impossible to associate it uniquely and in a
model-independent fashion with either one or the other electron population.
Based on measurements of the entire spectrum, the average strength of the magnetic
field in the x-ray production region is known to be in the tenths of mG, but it is
not known to high accuracy.
There are also large relative uncertainties in the radiation rates in some
regimes, particularly the highest energy.

However, we still can get a strong constraint on $c$ from the Crab nebula data.
This constraint, like the one derived in~\cite{ref-jacobson1}, is based upon the
fact that Lorentz violation may give rise to a maximum particle velocity. The
existence of electrons with large velocities then constrains the Lorentz-violating
parameters. For a particle moving in the direction of a unit vector $\vec{e}$,
the maximum allowed velocity is 
(to leading order in $c$) $1-c_{jk}e_{j}e_{k}-c_{0j}e_{j}$. We observe
via the Crab nebula synchrotron spectrum electrons with Lorentz factors as large
as $3\times10^{9}$. This means that the maximum velocity in the Crab-to-Earth
direction is greater than $1-6\times10^{-20}$, hence $c_{jk}e_{j}e_{k}+c_{0j}e_{j}
<6\times10^{-20}$. As in~\cite{ref-jacobson1}, this is a one-sided limit; one
sign of this combination of coefficients leads to a maximum velocity in the
relevant direction, but the other does not.

The direction $\vec{e}$ can be transformed into the standard sun-centered
celestial equatorial coordinates used in the study of Lorentz
violation~\cite{ref-bluhm4}. The location of the Crab
nebula is right ascension 5h 34m 32s, declination 22$^{\circ}$ 0$'$ 52$''$,
lying close to the ecliptic plane. So
the unit vector pointing from the nebula to the Earth has components
$e_{X}=-0.10$, $e_{Y}=-0.92$, and $e_{Z}=-0.37$.
This gives us the particular elements of the $c$ tensor that are constrained by
this measurement. The specific constraint is
\begin{eqnarray}
\left[0.01c_{XX}+0.85c_{YY}+0.14c_{ZZ}+0.09c_{(XY)}
+0.04c_{(XZ)}\right.& & \\
\left.+0.34c_{(YZ)}
-0.10c_{0X}-0.92c_{0Y}-0.37c_{0Z}\right]
& < & 6\times10^{-20}, \nonumber
\end{eqnarray}
where $c_{(jk)}$ is the symmetric sum $c_{jk}+c_{kj}$.
Similar constraints could be obtained for other well-resolved
synchrotron sources; this would provide further constraints on the
symmetric part of $c_{jk}$ and on the $c_{0j}$. [At leading order, the
antisymmetric part of $c_{jk}$ just represents a change in the representation of
the Dirac matrices, and it is already evident from (\ref{eq-vfull}) that it
will not contribute to the velocity.]


In order for these constraints to be valid, we must know that there are no
other effects that will interfere with our result. In particular,
we would like to address the question of whether quantum corrections would affect
the emission before Lorentz-violating corrections become important.
The leading order quantum corrections to the standard synchrotron
formulas may be found by making the replacement $\omega_{s}\rightarrow\omega_{s}
\left(1+\frac{\omega_{c}}{E}\right)$ in $\frac{1}{\omega_{s}}\frac{dI}
{d\omega_{s}}$~\cite{ref-schwinger2}. The corrections are negligible if
$\omega_{c}\ll E$, or equivalently if
$\gamma\ll\frac{m^{2}}{|e|B}$. This is $\gamma\ll \left(3\times 10^{13}\right)
B^{-1}$ if the particle is an electron and $B$ is measured in Gauss. If a typical
field strength is that within the Crab nebula, $B\sim 0.2$--$0.3$ mG, the maximum
values of $\gamma$ are extremely high. So
our classical treatment could apply up to scales well above those at which we
would expect to start seeing marked deviations from the conventional results.

Synchrotron radiation has already been used to set strong limits on
nonrenormalizable Lorentz-violating modifications of quantum electrodynamics.
Lorentz violation in synchrotron radiation is also theoretically interesting,
and there have been a number of prior analyses of the emission spectrum in
specific Lorentz-violating models.
In this paper, we have looked at the impact of the renormalizable SME
coefficient $c^{\nu\mu}$ on
synchrotron processes. Although we have used a number of standard
approximations to simply our analysis of the radiation, no approximations relating
to the Lorentz violation were required; the expressions (\ref{eq-v12sol}) and
(\ref{eq-v3sol}) are exact, valid to all orders in $c$. The $c^{\nu\mu}$
coefficients for electrons, particular the diagonal coefficients, can be difficult
to bound experimentally~\cite{ref-kost6,ref-bluhm4}. Since only a single fermion
is involved in synchrotron radiation, this is a process in which it is
relatively easy to isolate electron-specific effects, and the kind of constraints
we have obtained here should prove useful.


\section*{Acknowledgments}
The author is grateful to V. A. Kosteleck\'{y} for helpful discussions.
This work is supported in part by funds provided by the U. S.
Department of Energy (D.O.E.) under cooperative research agreement
DE-FG02-91ER40661.

\end{document}